\documentclass[utf8]{frontiersinFPHY_FAMS}

\setcitestyle{square} % for articles in the journals "Frontiers in Physics" and "Frontiers in Applied Mathematics and Statistics" 
\usepackage{url,hyperref,microtype}
\usepackage[onehalfspacing]{setspace}
\usepackage[left=1in, right=2.in]{geometry}
\usepackage{mathrsfs}
\usepackage{mathrsfs}
\usepackage{amsthm}
\usepackage{todonotes}

\makeatletter 
\@mparswitchfalse%
\makeatother
\normalmarginpar %for right-handed notes and lines, or 

\def\scri{\mathscr{I}}

\def\keyFont{\fontsize{8}{11}\helveticabold }
\def\firstAuthorLast{Panosso Macedo \& Zenginoglu} %use et al only if is more than 1 author
\def\Authors{Rodrigo Panosso Macedo\,$^{1,*}$, An\i l Zengino\u{g}lu\,$^{2}$}
\def\Address{$^{1}$1Niels Bohr International Academy, Niels Bohr Institute, Blegdamsvej 17, \\ 2100 Copenhagen, Denmark \\
$^{2}$Institute for Physical Science \& Technology, University of Maryland, \\ College Park, MD 20742-2431, USA}
\def\corrAuthor{Rodrigo Panosso Macedo}
\def\corrEmail{rodrigo.macedo@nbi.ku.dk}

\begin{document}
\onecolumn
\title[Hyperboloidal approach to QNMs]{Hyperboloidal Approach to Quasinormal Modes} 

\author[\firstAuthorLast ]{\Authors} %This field will be automatically populated
\address{} %This field will be automatically populated
\correspondance{} %This field will be automatically populated
\extraAuth{An\i l Zengino\u{g}lu \\ anil@umd.edu}
\maketitle

%%%%%%%%%%%%%%%%%%%%%%%%
\begin{abstract}
%%%%%%%%%%%%%%%%%%%%%%%%
Oscillations of black hole spacetimes exhibit divergent behavior near the bifurcation sphere and spatial infinity. In contrast, these oscillations remain regular when evaluated near the event horizon and null infinity. The hyperboloidal approach provides a natural framework to bridge these regions smoothly, resulting in a geometric regularization of time-harmonic oscillations, known as quasinormal modes (QNMs).

This review traces the development of the hyperboloidal approach to QNMs in asymptotically flat spacetimes, emphasizing both the physical motivation and recent advancements in the field. By providing a geometric perspective, the hyperboloidal approach offers an elegant framework for understanding black hole oscillations, with implications for improving numerical simulations, stability analysis, and the interpretation of gravitational wave signals.
 
\tiny
 \keyFont{ \section{Keywords:} hyperboloidal, frequency domain, quasinormal modes, black holes, non-selfadjoint operators} %All article types: you may provide up to 8 keywords; at least 5 are mandatory.
\end{abstract}

\pagebreak

%%%%%%%%%%%%%%%%%%%%%%%%
\section{Introduction}
%%%%%%%%%%%%%%%%%%%%%%%%

When a black hole (BH) spacetime is perturbed, gravitational waves (GW) carry the energy of the perturbation towards the BH horizon and to infinity. These perturbations show oscillations that decay exponentially at characteristic frequencies and are called quasinormal modes (QNM) \cite{kokkotas1999quasi, Nollert99_qnmReview, berti_quasinormal_2009, konoplya_quasinormal_2011}. Studying these QNMs is central to the black hole spectroscopy program~\cite{Dreyer:2003bv,Berti:2005ys,Berti:2016lat}, which aims to measure the oscillation frequencies from GW detections and thereby probe the BH geometry and its surrounding environment~\cite{kokkotas1999quasi,berti_quasinormal_2009,konoplya_quasinormal_2011,Franchini:2023eda}. The dominant quadrupole QNMs have already been measured in gravitational wave signals~\cite{LIGOScientific:2016aoc,LIGOScientific:2020tif,LIGOScientific:2021sio}, while the detection of higher modes remains under debate~\cite{Isi:2019aib,Capano:2020dix,Capano:2021etf,Cotesta:2022pci,Capano:2022zqm,Forteza:2022tgq,Finch:2022ynt,Abedi:2023kot,Carullo:2023gtf,Baibhav:2023clw,Nee:2023osy,Zhu:2023mzv,Siegel:2023lxl,Gennari:2023gmx}.

Mathematically, the QNM problem is often formulated as an eigenvalue problem, where QNM frequencies appear as the eigenvalues of a second-order differential operator. However, in their traditional representation, the corresponding QNM eigenfunctions grow exponentially near the black hole and at spatial infinity, which does not seem physically acceptable for small perturbations of a background spacetime \cite{schmidt_relativistic_1993}. Reformulating the problem using hyperboloidal surfaces---regular spacelike surfaces that extend smoothly from the black hole event horizon to null infinity---reveals that QNMs are globally regular \cite{zenginoglu_geometric_2011, ansorg_spectral_2016}. This geometric regularization\footnote{\label{fn:refs_qnm}Note that we focus here on the geometric developments around the hyperboloidal framework in asymptotically flat spacetimes. The analytic aspects of QNM regularity beyond the mere coordinate singularity of standard time slices were clarified in a series of papers \cite{horowitz2000quasinormal, vasy2013microlocal, warnick_quasinormal_2015, Gajic:2019oem, Gajic:2024xrn}, discussed in Sec.~\ref{sec:analysis}.} of time-harmonic black hole perturbations has found many recent applications, which we review in this paper.

%%%%%%%%%%%%%%%%%%%%%%%%
\section{The traditional approach to QNMs}\label{sec:traditional}
%%%%%%%%%%%%%%%%%%%%%%%%

The Schwarzschild solution, the simplest black-hole (BH) solution to Einstein's equations, is given by
\[ ds^2 = - f(r) dt^2 + \frac{1}{f(r)} dr^2 + r^2 d\varpi^2, \quad \textrm{with} \quad f(r) = 1-\frac{2M}{r},\]
where $d\varpi^2 = d\theta^2 + \sin^2\theta d\phi^2$ is the metric on the unit sphere, and $M$ the black-hole's mass. Perturbations of this solution are described by the Regge-Wheeler-Zerilli type wave equation:
\begin{equation} 
\label{eq:wave}
    \left(-\partial_t^2 + \partial_{r_\ast}^2 - V(r_\ast)\right) u(t,r_\ast) = 0,
\end{equation}
where $r_\ast\in(-\infty,\infty)$ is the tortoise coordinate and $V(r_\ast)$ is the effective potential that behaves as $ V\sim f(r)$ near the black-hole horizon and as $V\sim 1/r^2$ toward spatial infinity. 

Solutions to \eqref{eq:wave} evolve through a transient phase, followed by a ringdown characterized by exponentially damped vibrations (QNMs) \cite{VishuNature70}, and eventually a polynomial, non-oscillatory decay known as the tail \cite{Price72,Gundlach_94}.

To analyze the QNM phase, one typically considers time-harmonic solutions
\begin{equation}
\label{eqn:time_harmonic} 
u(t,r_\ast) = e^{-i\omega t} R(r_\ast),
\end{equation}
that reduce the wave equation to a Helmholtz equation,
\begin{equation} 
\label{eqn:helmholtz} 
\left(\frac{d^2}{d r_\ast^2} + \omega^2 - V(r_\ast)\right) R(r_\ast) = 0.
\end{equation}
Sommerfeld recognized in 1912 that the Helmholtz equation, in stark contrast to the elliptic case, does not admit unique solutions even when we require that the solution vanishes at infinity \cite{sommerfeld1912greensche, schot_eighty_1992}. To ensure uniqueness, an outgoing radiation condition must be imposed. In the BH context, a Sommerfeld condition applies also near the BH. We therefore impose
\begin{equation}
\label{eqn:sommerfeld} 
\lim_{r_\ast \to \pm \infty} \left(\frac{d}{d{r_\ast}} \mp i\omega\right) R(r_\ast) = 0 \quad \Leftrightarrow \quad R(r_\ast) \sim e^{\pm i\omega r_\ast} \ \textrm{as} \ r_\ast\to \pm \infty. 
\end{equation}
% Curiously, we must impose a boundary condition on the time-harmonic wave equation by hand to ensure the uniqueness of solutions. The underlying reason is that the original wave equation \eqref{eq:wave} is time-symmetric. It admits solutions that are incoming from infinity\footnote{Sommerfeld considered waves that are incoming from infinity as "physically meaningless." We now understand that such incoming waves provide an idealization for a scattering problem \cite{compere2023asymptotic, kehrberger2024case}} and outgoing toward infinity. Depending on which class of solutions one is interested in, one must impose the corresponding boundary condition by hand.

It turns out, however, that the boundary conditions \eqref{eqn:sommerfeld} are not sufficient \cite{Nollert99_qnmReview, Nollert92} and a more precise notion of purely outgoing solution is needed to uniquely define the QNMs \cite{bachelot1993resonances}. 
% These property is usually underappreciated by the community, but its manifestation becomes evident in the hyperboloidal formalism as we review in sec.~\ref{sec:analysis}.
The formal definition of QNMs followed a different route than the intuitive notion of QNMs as the eigenvalues of a given differential operator.  

The time-harmonic Ansatz \eqref{eqn:time_harmonic}, closely related to a Fourier transformation, provides a general formalism oblivious to the specific form of initial data causing the perturbation. To define QNMs formally, one considers an initial value problem. Then, a Laplace transformation\cite{Nollert99_qnmReview, Leaver86, SunPrice88,Nollert92} leads to a inhomogeneous spatial differential equation, with a source term accounting for the initial data.
% Nonetheless, the homogenous eq.~\eqref{eqn:helmholtz} still plays an important role in the formal definition of QNM, as its solutions provide the building blocks for the Green's function and Wronskian determinant associated with the wave operator. 
One must then ensure that the spacetime solution $u(t,r_*)$ remains bounded as $t\rightarrow \infty$~\cite{KayWald87}. In the Laplace formalism, the solution $u(t,r_*)$ results from the convolution of the Green's functions with the source term carrying information from the initial data. 
The inverse Laplace transformation requires an integration along a frequency values $\omega_I > 0$, and there is only one possible choice of homogenous solutions $R_\pm(r_*;\omega)$ to construct the correct Green's function: they must satisfy the boundary conditions \eqref{eqn:sommerfeld} at both ends $r_* \rightarrow \pm \infty$.
% Specifically, $R_+(r_*;\omega)$ must satisfy the boundary condition \eqref{eqn:sommerfeld} as $r_* \rightarrow \infty$ (no ingoing radiation from past null infinity) and whereas $R_-(r_*;\omega)$ must satisfy eq.~\eqref{eqn:sommerfeld} as $r_* \rightarrow -\infty$ (no outgoing radiation from the white hole horizon)\footnote{These solutions are, respectively, also often referred in the literature as {\em up} and \em in solutions, e.g.~\cite{Pound:2021qin}.}\cite{Nollert99_qnmReview,kokkotas1999quasi,Nollert92}. 
Once the Green's functions are fixed in the complex half-plane $\omega_I > 0$, one analytically extends them into the region $\omega_I < 0$. The QNMs are then uniquely defined as the poles of Green's functions, or equivalently in the one-dimensional case, the roots of the Wronskian. 
% The condition $W(\omega_n) = 0$ implies that $R_+ \propto R_-$, and therefore $R_\pm$ satisfy the boundary conditions \eqref{eqn:sommerfeld} at both ends $r_* \rightarrow \pm \infty$.

The Laplace approach uniquely defines QNMs via Green's functions, bypassing the notions of eigenvalues and eigenfunctions. Such definition via this Green's functions is also understood under the Lax-Phillips approach\cite{zworski2017mathematical,dyatlov2019mathematical}. However, this definition still allows QNM functions to blow up asymptotically, creating a puzzle: while black hole stability demands that linearized perturbations decay over time, the associated time-harmonic perturbations remain singular in the asymptotic regions.

The resolution lies in the global structure of spacetime. The QNM behavior at asymptotic boundaries results from the singular properties of the coordinates used in eq.~\eqref{eqn:helmholtz}. In Schwarzschild coordinates, as $r_* \rightarrow \pm \infty$, the limits correspond to spatial infinity $i^0$ and the bifurcation sphere $\mathcal{B}$. These loci connect to future and past null infinity at $i^0$ and white and black hole horizon at $\mathcal{B}$, and the blow-up of QNM eigenfunctions is a coordinate effect due to the accumulation of infinitely many time surfaces thereon.
%at $i^0$ and $\mathcal{B}$.

When QNMs are represented on regular, hyperboloidal time slices, they do not exhibit this unbounded growth\footref{fn:refs_qnm} \cite{zenginoglu_geometric_2011, ansorg_spectral_2016,Gajic:2024xrn}, as we discuss in the next section.

\section{The hyperboloidal approach to QNMs}
%%%%%%%%%%%%%%%%%%%%%%%%

The singularity of Schwarzschild time slices at the bifurcation sphere is well-known today, but understanding its causal structure took over four decades \cite{israel1987dark, senovilla_1965_2015, nielsen2022origin, landsman_penroses_2022}. Given this singularity, it is not surprising that QNMs blow up near the black hole, but they also blow up near spatial infinity. Thus, switching to regular coordinates at the bifurcation sphere doesn't resolve the issue.

Part of the historical confusion about BHs was that it takes infinite Schwarzschild time for radiation to fall into a BH. The same statement is true concerning spatial infinity: it takes infinite Schwarzschild time for outgoing radiation to reach spatial infinity. Because this is ``reasonable" from a physical point of view, it has been widely accepted that QNMs have a singular representation at both asymptotic regions.

The first suggestion that outgoing perturbations are regular in the frequency domain toward null infinity was made by Friedman and Schutz in a 1975 paper on the stability of relativistic stars \cite{friedman_stability_1975}. Friedman and Schutz recognize the problem with standard time slices where outgoing modes behave asymptotically like $r^k e^{i\omega (t-r)}$ implying that stable modes with $\omega_I < 0$ grow exponentially as $r\to\infty$. To make the representation finite, they recommend to use null hypersurfaces. In a footnote, they comment that the representation is regular also ``if one uses a spacelike hypersurface that is only asymptotically null." 

Schmidt picked up this idea in a 1993-essay for the  Gravity Research Foundation on relativistic stellar oscillations \cite{schmidt_relativistic_1993} arguing that QNMs on hyperboloids ``are represented by proper eigenvalues and eigenfunctions." However, the presentation includes no details beyond the 1975 paper.  

To understand why it took almost 20 years from Schmidt's essay \cite{schmidt_relativistic_1993} to the construction of a regular geometric framework to describe QNMs in asymptotically flat spacetimes\footref{fn:refs_qnm} \cite{zenginoglu_geometric_2011, ansorg_spectral_2016,Gajic:2024xrn}, we provide a short historical review of hyperboloidal coordinates.

%%%%%%%%%%%%%%%%%%%%%%%%
\subsection{A brief history of hyperboloidal coordinates}

The central role of spacetime hyperbolas in relativity was recognized already by Minkowski in 1908 \cite{minkowski_1908}. The Milne model from 1933 \cite{milne1936relativity} or Dirac's point-form of quantum field theory from 1949 are hyperboloidal \cite{dirac1949forms}. In the 1970s, hyperboloidal studies were performed for the analysis of wave equations \cite{chen_solutions_1970, chen_hyperboloidal_1971, strichartz_harmonic_1973} and quantum field theory \cite{fubini1973new, disessa1974quantization, sommerfield_quantization_1974, hostler1978coulomb}. However, these early studies use a time-dependent formulation in which time freezes at null infinity.

The first hyperboloidal coordinates foliating null infinity are implicit in Penrose's work on the global causal structure of spacetimes via conformal compactification \cite{penrose_zero_1965, penrose_asymptotic_1963}. Indeed, one can obtain a hyperboloidal surface from any textbook discussing the Penrose diagram simply by looking at the level sets of Penrose time~\cite{zenginouglu2024hyperbolic}. In the context of numerical relativity, it was recognized that hyperboloidal time functions that asymptotically approach the retarded time should be beneficial for the computation of gravitational waves \cite{smarr_kinematical_1978, eardley_time_1979}. Explicit hyperboloidal coordinates in black-hole spacetimes were constructed in the context of the analysis of constant mean curvature foliations \cite{brill_k_1980}. A remarkable but largely ignored paper by Gowdy in 1981 includes many key elements of the hyperboloidal approach used today in black-hole perturbation theory \cite{gowdy_wave_1981}, including the height function approach to preserve time-translation symmetry, compactification fixing null infinity (scri-fixing), hyperboloidal solutions to the wave equation, and the structure of time-harmonic solutions relevant for the frequency domain. These ideas were not picked up by the community at the time. 

The first systematic study of the hyperboloidal initial value problem for Einstein equations was initiated by Friedrich in 1983 \cite{friedrich1983cauchy}. Friedrich devised a reformulation of the Einstein equations with respect to a conformally rescaled metric that is regular across null infinity. The conformal field equations are well-suited for the analysis of the asymptotic behavior of Einstein's equations and have led to seminal results such as the nonlinear, semi-global stability of de Sitter-type and Minkowski-type spacetimes \cite{friedrich1986existence, friedrich1986ngeo}. The developments around conformal field equations and attempts to use them numerically are reviewed in \cite{frauendiener2004conformal,Kroon2016}.

% When Schmidt published his essay suggesting the application of hyperboloidal coordinates to describe QNMs in 1993, hyperboloidal surfaces and their asymptotic behavior were well-known among mathematical relativists. It is curious that even though many abstract results were proven, an explicit construction of hyperboloidal coordinates in black-hole spacetimes suitable for perturbation theory was not available.

Twenty years after Gowdy's paper, Moncrief presented the hyperboloidal compactification of Minkowski spacetime using time-shifted hyperboloids in an unpublished talk \cite{Moncrief00} leading to the first numerical studies using hyperboloidal foliations in Minkowski spacetime \cite{husa2003numerical, FodorRacz04, fodor2008numerical, bizon2009universality}. Around this time, various suggestions for hyperboloidal coordinates and numerical simulations in black-hole spacetimes were made \cite{gentle_constant_2001, gowdy_compactification_2001, schmidt_data_2002, pareja_constant_2006, calabrese_asymptotically_2006, misner2006excising}. 

The construction widely used today in black-hole perturbation theory is based on scri-fixing coordinates with time-shifted hyperboloids presented in 2008 \cite{zenginoglu_hyperboloidal_2008}. The idea is to combine the height function technique that preserves the time-symmetry of the underlying spacetime with an explicit radial compactification whose singular Jacobian at infinity is proportional to a prescribed conformal factor. In the following years, this method was used primarily in the time domain for solving wave-propagation problems \cite{zenginoglu_tail_2008, zenginouglu2009gravitational, zenginoglu_spacelike_2009, zenginoglu_asymptotics_2010, zenginouglu2010hyperboloidal, bizon_saddle-point_2010, lora2010evolution, jasiulek2011hyperboloidal, racz2011numerical, vega2011effective, bernuzzi_binary_2011, bernuzzi_layer_2011, zenginouglu2011null}. 

The translation of the hyperboloidal method to the frequency domain was presented in \cite{zenginoglu_geometric_2011}, where it was demonstrated that hyperboloidal time functions regularize the QNM eigenfunctions in the asymptotic domains. Warnick used a related idea in \cite{warnick_quasinormal_2015} for AdS spacetimes in which spatial slices are naturally hyperbolic (see also \cite{horowitz2000quasinormal, vasy2013microlocal}). The first detailed analysis of QNMs in asymptotically flat black-hole spacetimes using the hyperboloidal approach was presented in \cite{ansorg_spectral_2016}. We summarize the basic ideas of the hyperboloidal approach below.

%%%%%%%%%%%%%%%%%%%%%%%%
\subsection{A geometric framework}

% The asymptotic behavior of the QNM eigenfunctions is different toward the BH and toward infinity. These two conditions cannot be satisfied simultaneously. The only way both conditions can be satisfied \textbf{simultaneously} is if the hypersurface connects the horizon with null infinity in a spacelike manner. Such surfaces are horizon-penetrating-hyperboloidal.

\begin{figure}[th!]
    \centering
    \includegraphics[width=\columnwidth]{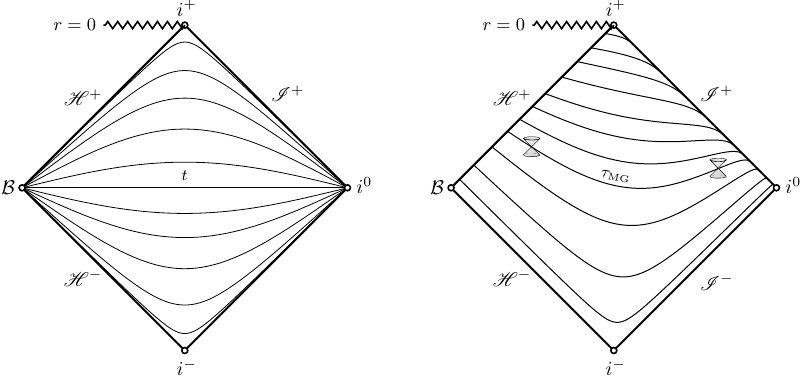}	
    \caption{Penrose diagrams of the exterior domain in Schwarzschild spacetime contrasts the level sets of the standard Schwarzschild time (left panel) and the hyperboloidal minimal gauge (right panel). Schwarzschild time slices intersect at the bifurcation sphere, $\mathcal{B}$, and spatial infinity, $i^0$. Minimal gauge slices provide a smooth foliation of the future event horizon, $\mathcal{H^+}$, and future null infinity, $\scri^+$.}
    \label{fig:penrose_diagram}
    \end{figure}

The construction of globally regular coordinates consists of a time transformation respecting the time symmetry of the background, a suitable spatial compactification, and conformal rescaling \cite{zenginoglu_hyperboloidal_2008}. We first introduce the time function $\tau$ via \cite{gowdy_wave_1981, calabrese_asymptotically_2006, zenginoglu_hyperboloidal_2008}
\begin{equation} \label{eqn:tau} \tau = t + h(r). \end{equation} 
The time transformation implies an exponential scaling in frequency domain \cite{zenginoglu_geometric_2011}. Writing the time-harmonic ansatz in \eqref{eqn:time_harmonic} with respect to the new time coordinate in \eqref{eqn:tau}, we get
\[ u(t,r_\ast) = e^{-i\omega t} R(r_\ast) = e^{-i \omega \tau} e^{i \omega h}R(r_\ast) = e^{-i \omega \tau}\bar{R}(r_\ast) . \]
The rescaled radial function $\bar{R}(r_\ast) = e^{i \omega h} R(r_\ast)$ is regular both near the event horizon and toward null infinity. To see this in an explicit example, consider the height function for the so-called {\em minimal gauge} \cite{schinkel_initial_2014,PanossoMacedo:2018hab,PanossoMacedo:2023qzp}
\[
 h_{\textrm{MG}}(r) = -r + 2M\log\left| \dfrac{r}{2M}-1 \right| - 4M \log\left(\dfrac{r}{2M} \right) = - r_\ast - 4M \log\left(\dfrac{r}{2M}\right). 
 \]
The minimal gauge height function has the following asymptotic behavior
\[ h_{\textrm{MG}} \sim -r_\ast \ \mathrm{for} \ r\to\infty, \qquad  h_{\textrm{MG}} \sim +r_\ast \ \mathrm{for} \ r\to 2M. \]
The height function regularizes the QNM eigenfunctions in the asymptotic domains. The regularity of the QNM eigenfunctions is directly related to the regularity of the minimal gauge at the asymptotic boundaries near the horizon and near infinity (see Fig.\ref{fig:penrose_diagram}). The minimal gauge is unique in its simplicity and appears in different setups as a natural construction \cite{aimer_jack_quasinormal_2023,Ripley:2022ypi}. Surprisingly, the minimal gauge was implicitly used by Leaver in his papers on QNMs in BH spacetime \cite{leaver1985analytic,leaver1990_RN}. Related hyperboloidal regularization procedures have been suggested over the years by various authors without an explicit recognition of the geometric background of their construction \cite{dolan_expansion_2009, dolan_quasinormal_2010,jansen_overdamped_2017, langlois_black_2021, langlois_asymptotics_2021,chung_spectral_2023, chung_kerr_2023}.

In \cite{zenginoglu_geometric_2011}, it was shown that the time translation must be combined with a suitable rescaling to arrive at a regular representation of QNMs. The rescaling takes into account the asymptotic fall-off behavior of the QNM eigenfunctions toward the BH and toward infinity. The resulting equations have short-range potentials suitable for compactification of the exterior black region from the radial coordinate $r\in [r_h, \infty)$  --- or equivalently, $r_* \in (-\infty, +\infty$) ---  into a compact domain $\sigma \in [\sigma_h, \sigma_{\scri^+}]$. This rescaling is related to the conformal compactification of black-hole spacetimes. 

The external boundary conditions \eqref{eqn:sommerfeld} are automatically satisfied in terms of a radially compact hyperboloidal coordinates $(\tau,\sigma)$\cite{zenginoglu_geometric_2011,PanossoMacedo:2023qzp}, when the underlying function $\overline R(\sigma)$ is regular at the black-hole horizon ($\sigma_h$) and future null infinity ($\sigma_{\scri^+}$). Thus, we no longer need to impose boundary conditions to the wave equation by hand. The boundary condition is replaced by a regularity condition on the underlying solution $\overline R(\sigma)$ in the entire domain $\sigma \in [\sigma_h, \sigma_{\scri^+}]$. In practical terms, one derives the regularity condition at the boundary directly from the hyperboloidal differential equation. When formulating the frequency-domain problem in coordinates $(\tau,\sigma)$, the resulting differential equation equivalent to Helmoltz equation~\eqref{eqn:helmholtz} assumes a generic form\cite{ansorg_spectral_2016, PanossoMacedo:2018hab, PanossoMacedo:2023qzp}
\begin{equation}
\label{eq:hyp_ODE}
\Bigg( \alpha_2(\sigma)\dfrac{d^2}{d\sigma^2} +  \alpha_1(\sigma)\dfrac{d}{d\sigma} +  \alpha_0(\sigma) \Bigg) \overline R(\sigma) = 0,
\end{equation}
with coefficients $\alpha_2$, $\alpha_1$ and $\alpha_0$ depending on the particular choice for the hyperboloidal height function. The most important property of the above equation is that it is a {\em singular} ordinary differential equation, i.e, its principal part behaves as $\alpha_2 \sim (\sigma - \sigma_{\scri^+})^2(\sigma - \sigma_h)$ and therefore $\alpha_2(\sigma_h)=\alpha_2(\sigma_{\scri^+})=0$. Hence, at the boundaries, eq.~\eqref{eq:hyp_ODE} provides us directly with the relation between the field and its first $\sigma$-derivatives serving as boundary data ensuring a bounded solution. From the spacetime perspective, and the resulting wave equation, the same condition $\alpha_2(\sigma_h)=\alpha_2(\sigma_{\scri^+})=0$ ensures that, at the boundaries, the light cones point outwards the numerical domain, or equivalently, that the characteristic speeds of incoming modes vanish \cite{zenginoglu_geometric_2011,PanossoMacedo:2023qzp}.

The finite behaviour of the function $\overline R(\sigma)$ is one of the most important aspects in the hyperboloidal approach. As we discuss below, this feature allows us to unveil new properties of the QNM eigenfunctions, develop novel numerical algorithms and attack new problems relevant to black-hole physics and gravitational wave astronomy.

%\todo{Distinguish between boundary conditions and data}

% \missingfigure{Make a figure demonstrating the physical reason for the singularity of the QNM eigenfunctions with null directions.}

\subsection{From geometry to analysis}\label{sec:analysis}

The hyperboloidal framework regularizes solutions to the Helmholtz problem \eqref{eqn:helmholtz}. In fact, {\em any} bounded solution satisfying the singular ordinary differential equation \eqref{eq:hyp_ODE} automatically fulfills the Sommerfeld conditions \eqref{eqn:sommerfeld}. One would naively think that bounded solutions exist only at the QNMs frequencies. However, we saw in Sec.~\ref{sec:traditional} that the conditions \eqref{eqn:sommerfeld} are {\em necessary, but not sufficient} to specify the QNM problem uniquely. In the traditional formulation, the QNM eigenfunctions grow asymptotically. The complex plane spanned by the frequency $\omega$ might contain regions with solution satisfying \eqref{eqn:sommerfeld}, but contaminated by unwanted solutions that decrease at the boundaries. Removing the asymptotic blow-up allows us to peek directly into these unphysical solutions, which exist in the entire half-plane ${\rm Im}(\omega)<0$~\cite{warnick_quasinormal_2015,ansorg_spectral_2016}.

\begin{figure}[th!]
\includegraphics[width=\columnwidth]{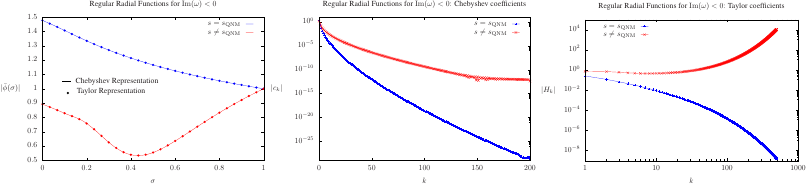}			
    \caption{Solutions to hyperboloidal radial equation (left panel). bounded solutions exist in the entire half-plane for ${\rm Im}(\omega)<0$, regardless whether $\omega$ is a QNM or not. The continuous lines were obtained with a spectral method code based on a Chebyshev representation of the solution. Independently, the dots result from a Taylor representation of the solution as power series around the horizon (Leaver's strategy).  The QNM eigenfunctions are characterised by the solutions with a higher degree of regularity, heuristically verified via the faster decay of the corresponding Chebyshev coefficients $|c_k|$  (middle panel) or the the asymptotic decay of the Taylor coefficients $|a_k|$. }
    \label{fig:reg_solutions}
\end{figure}

The left panel of Fig.~\ref{fig:reg_solutions} shows two solutions to Eq.~\eqref{eq:hyp_ODE} which are bounded in the entire exterior BH domain, from $\sigma=0$ representing future null infinity and $\sigma=1$ the BH horizon: the solution in blue is obtained at a given QNM frequency $\omega_{\rm QNM}$, whereas the solution in red corresponds to a given frequency in the half-plane ${\rm Im}(\omega)<0$, but with $\omega\neq \omega_{\rm QNM}$. These solutions are obtained with two different numerical approaches: the solid line results from solving the ODE with a Chebyshev collocation point spectral method\cite{Jaramillo:2020tuu}, whereas the dotted points arise when using Leaver's Taylor expansion, which corresponds to a frequency domain hyperboloidal formulation in the minimal gauge\cite{ansorg_spectral_2016,PanossoMacedo:2018hab,PanossoMacedo:2019npm}. Both strategies yield the same results. At first glance, there is nothing special in the behavior of the solutions that allows us to distinguish a QNM from a non-QNM eigenfunction. Indeed, it is even possible to specify a hyperboloidal initial data such that the corresponding time evolution has an arbitrary exponentially damped oscillation~\cite{ansorg_spectral_2016}.

What distinguishes a QNM from a non-QNM solution is their regularity class. By studying the convergence rate of their discrete numerical representation, one can infer that these functions belong to different regularity classes. The middle panel of Fig.~\ref{fig:reg_solutions} shows the Chebyshev coefficients from the Chebyshev collocation point spectral method. These coefficients decay exponentially for $C^{\Omega}$ analytic functions or algebraically for $C^k$ singular functions. We observe an intermediary decay, suggesting the regularity class of these functions is between $C^k$ and $C^{\Omega}$. The Chebyshev coefficients for QNM eigenfunctions decay faster than for non-QNM eigenfunctions, indicating QNM eigenfunctions belong to a better regularity class. A similar conclusion arises from the Taylor expansion coefficients (right panel ~\ref{fig:reg_solutions}). For QNM eigenfunctions, $|a_k|$ decays asymptotically. For non-QNM functions, $|a_k|$ grows asymptotically. Even though the series does not converge absolutely, it converges conditionally due to oscillations in ${\rm Im}(a_k)$ (left panel of Fig.~\ref{fig:reg_solutions}). These conclusions are formalized by interpreting QNM as a formal eigenvalue problem of the generator of time translations for a null foliation, acting on an appropriate Hilbert space \cite{horowitz2000quasinormal, vasy2013microlocal, warnick_quasinormal_2015, Gajic:2019oem, Gajic:2019qdd, bizon_toy_2020, warnick_completeness_2022,Gajic:2024xrn}, where QNM eigenfunctions belong to the Gevrey-2 regularity class.

%%%%%%%%%%%%%%%%%%%%%%%%
\section{Applications}
%%%%%%%%%%%%%%%%%%%%%%%%
The QNM problem plays a fundamental role in the era of gravitational wave astronomy. The BH spectroscopy program faces three main challenges: (i) the measurability of the QNMs frequencies, limited by the GW detection signal-to-noise ratios; (ii) the relevance of nonlinear effects to the ringdown dynamics; and (iii) the QNMs spectral instability. As discussed in the previous sections, hyperboloidal formalism provides crucial theoretical tools to tackle different aspects of these challenges. 
% In fact, the applicability of the hyperboloidal formalism extends beyond the ringdown phase and includes, for example, the modeling of extreme mass ratio inspirals via the gravitational self-force programme\cite{wardell2014self,Pound:2021qin,PanossoMacedo:2022fdi,DaSilva:2023xif,PanossoMacedo:2024pox}. This section summarises key applications and breakthroughs the hyperboloidal framework introduced to the field.   

\subsection{QNM Excitation factors and tail decay}\label{sec:excitation}
Even though challenge (i) mainly concerns the GW detection's signal-to-noise ratio, it heavily relies on accurate predictions for the expected QNM excitations \cite{Carullo:2024smg}. 
% Recall that the oscillation and decaying time scales specified by the QNMs depend solely on the wave equation. However, 
The excitation of each QNM depends on the particular initial perturbation triggering the dynamics. This perturbation also excites the late-time power-law tail decay. Determining these excitation factors has always been challenging due to the blow-up of the underlying modes at the bifurcation sphere and spatial infinity~\cite{Berti:2006wq}. A common approach to avoid the infinities at the bifurcation sphere when calculating integrals along the physical coordinate $r_* \in (-\infty, \infty)$ is to deform the integration path into the complex plane~\cite{Zimmerman:2014aha,Green:2022htq,London:2023aeo}.

The hyperboloidal formalism offers an alternative strategy to determine such excitation factors due to the globally regular behavior of the QNM eigenfunctions. The direct identification of Leaver's continued fraction strategy with spacetime solutions defined on hyperboloidal hypersurfaces allows the further development of the Leaver method to calculate QNMs (and tail decay) excitation factors for problems formulated on hyperboloidal slices~\cite{ansorg_spectral_2016,PanossoMacedo:2018hab}. While Leaver's method relies on a Taylor expansion around the horizon for the underlying hyperboloidal functions, the strategy can be adapted to directly solve a linear partial differential equation having the QNM excitation amplitude as an unknown parameter in the equation \cite{Ammon:2016fru}, or alternatively via the so-called Keldysh scheme \cite{Besson:2024adi}. The hyperboloidal formalism is also essential for recent advances in the understanding of the role played by the tail decay in BH spectroscopy\cite{Zenginoglu:2012us,Cardoso:2024jme,DeAmicis:2024not}.

\subsection{Quadratic QNMs}\label{sec:quadratic}
Since GR is a nonlinear theory, challenge (ii) emphasizes that BH spectroscopy must also account for second-order, quadratic perturbations~\cite{London:2014cma,Cheung:2022rbm, Mitman:2022qdl, Zlochower:2003yh,Sberna:2021eui,Baibhav:2023clw,Redondo-Yuste:2023seq,Cheung:2023vki,ma2024excitation,Zhu:2024rej,Bucciotti:2024zyp}. The quadratic coupling of first-order solutions dictates the dynamics at second order in perturbation theory\cite{Brizuela:2009qd,Loutrel:2020wbw,Spiers2023}. When formulated in the standard $t$ slices, the blow-up of QNM eigenfunctions at the bifurcation sphere and spatial infinity imposes severe restrictions for second-order studies, both at theoretical and numerical levels. The hyperboloidal framework for black-hole perturbations beyond the linear order becomes indispensable for regular evolutions \cite{Zhu:2023mzv,Redondo-Yuste:2023seq,May:2024rrg}, as well as for studies in the frequency domain \cite{ma2024excitation,Bourg:2024jme}.

\subsection{QNM instability and the pseudospectrum}\label{sec:pseudospectrum}
Apart from his groundbreaking work in QNM \cite{VishuNature70}, Vishveshwara also highlighted that the QNM spectra is very sensitive to small modifications in the black-hole potential\cite{Vishveshwara:1996jgz,Aguirregabiria:1996zy}. At the same time, the QNM spectra destabilisation was also observed by Nollter and Price\cite{Nollert:1996rf,Nollert:1998ys}, but the phenomenon's impact in the BH spectroscopy programme has been largely overlooked over the past decades. Only recently has the challenge (iii) gained a greater attention\cite{Jaramillo:2020tuu,Jaramillo:2021tmt,Gasperin:2021kfv,Cheung:2021bol,alsheikh:tel-04116011,Berti:2022xfj,Jaramillo:2022kuv,Cardoso:2024mrw,Hirano:2024fgp,warnick2024stability,boyanov2024destabilising}. 

Small modifications in oscillatory frequencies for wave equations result in minor spectral responses only if the wave operators are self-adjoint. However, the flow of GWs into the BH and out into the wave zone places BH perturbation theory within the framework of non-self-adjoint operators. The successful application of non-self-adjoint operator theory to gravitational systems was only made possible by the hyperboloidal approach to black-hole perturbations\cite{Jaramillo:2020tuu} (see also ref.~\cite{galkowski2020outgoing} for an alternative approach akin to ``complex scaling"). In this approach, one can use the mathematical formalism of pseudospectra\cite{trefethen2005spectra} to study the QNM spectral instability\cite{Jaramillo:2020tuu} and perform a non-modal analysis\cite{Jaramillo:2022kuv} that a traditional mode analysis might overlook. Since the breakthrough offered by the hyperboloidal framework, the analysis of QNM pseudospectra has been performed in several different contexts, from astrophysically relevant scenarios to applications in the gauge-gravity duality\cite{Destounis:2021lum,Boyanov:2022ark,Arean:2023ejh,Sarkar:2023rhp,Destounis:2023nmb,Boyanov:2023qqf,Cownden:2023dam,Chen:2024mon,Carballo:2024kbk,Garcia-Farina:2024pdd}

%%%%%%%%%%%%%%%%%%%%%%%%
\section{Discussion}
%%%%%%%%%%%%%%%%%%%%%%%%

The hyperboloidal approach to QNMs offers a geometric regularization of black-hole perturbations. By connecting the regular oscillations near BHs with those observed far away, this method bypasses the problematic divergences inherent in the traditional approach at the bifurcation sphere and spatial infinity.

With hindsight, the hyperboloidal approach relies on a simple coordinate transformation that resolves the asymptotic singularity of the standard time \cite{zenginoglu_hyperboloidal_2008}. It is astonishing that it took decades for relativists to adopt regular coordinates to describe black-hole perturbations. We suspect that part of the confusion arose from the asymptotic behavior of time functions. It is not widely appreciated that the standard time coordinate in flat spacetime is singular at infinity with respect to the causal structure. Large-scale wave phenomena demand coordinates tailored to wave propagation—characteristic or hyperboloidal. This approach is critical not only for gravitational waves but also for addressing general wave propagation problems.

In recent years, the hyperboloidal approach has led to significant breakthroughs in the study of black-hole perturbations. As discussed, the regularity of the QNM eigenfunctions in the frequency domain enables a direct identification of the QNM excitation factors and tail decay \ref{sec:excitation}, facilitates the efficient computation of second-order perturbations \ref{sec:quadratic}, and supports the analysis of the QNM pseudospectrum \ref{sec:pseudospectrum}. Moreover, recent work has demonstrated that the hyperboloidal method can be extended to non-relativistic operators \cite{burgess2024hyperboloidal}, further broadening its scope and applicability.

From a numerical perspective, finding the optimal choices among the many ways to construct hyperboloidal coordinates, particularly for high-precision and large-scale simulations, remains a challenge. Exploring gauge conditions and optimizing numerical algorithms to leverage advanced computational resources will be essential for practical applications, going beyond linear perturbations and including the numerical solution of the full Einstein equations along hyperboloidal surfaces \cite{peterson2024spherical, vano2024height}. 

Much of the current work has focused on asymptotically flat, vacuum spacetimes. The formalism for black hole perturbation theory is fully developed for spherically symmetric spacetimes, but the same concepts are also valid for the Kerr solution\cite{PanossoMacedo:2019npm,Ripley:2022ypi}. The hyperboloidal approach is versatile and extendable to more general settings, including those with different asymptotic structures, and nonvaccum spacetimes. Developing these extensions will be crucial for applying this framework to a broader range of physical scenarios.

%%%%%%%%%%%%%%%%%%%%%%%%
\section*{Funding}
%%%%%%%%%%%%%%%%%%%%%%%%
RPM acknowledges support from the Villum Investigator program supported by the VILLUM Foundation (grant no. VIL37766) and the DNRF Chair program (grant no. DNRF162) by the Danish National Research Foundation and the European Union’s Horizon 2020 research and innovation programme under the Marie Sklodowska-Curie grant agreement No 101131233. 
AZ is supported by the National Science Foundation under Grant No.~2309084.

\bibliographystyle{Frontiers-Vancouver}
\bibliography{refs}

\end{document}